\documentclass[superscriptaddress,aps,pre,twocolumn]{revtex4-1}
\usepackage[utf8]{inputenc}
\usepackage{amsmath}
\usepackage{amssymb}
\usepackage{amsthm}
\usepackage{thmtools}
\usepackage{refcount}
\usepackage{siunitx}
\usepackage{graphicx}
\usepackage{hyperref}
\hypersetup{colorlinks,allcolors=black}
\usepackage{xcolor}

\usepackage{natbib}
 \usepackage[capitalise]{cleveref}
\usepackage{microtype}
\usepackage{bm}
\usepackage{lipsum}
\usepackage[english]{babel}
\usepackage{color}
\usepackage{graphicx}
\usepackage{xcolor}
\usepackage{braket}

\newcommand{\bp}{\mathbf{p}}

\begin{document}
\author{Haidar Al-Naseri}
\email{hnaseri@stanford.edu}
\affiliation{Stanford PULSE Institute, SLAC National Accelerator Laboratory, Menlo Park, California 94025, USA}

\title{Electron-positron pair annihilation in kinetic plasma}

\begin{abstract}
The process of electron-positron pair annihilation, driven by strong fields (Inverse Schwinger mechanism) and high-frequency waves, is studied using the Dirac–Heisenberg–Wigner formalism. In an electron-positron plasma, the presence of a strong field leads to both pair creation and annihilation. Depending on plasma properties such as non-degeneracy and the momentum distribution, pair annihilation can dominate over pair creation. The energy released from annihilated pairs can lead to an enhancement of the field energy, provided that the plasma effectively blocks the creation of new pairs. Additionally, pair annihilation induced by high-frequency waves is shown to occur when the photon energy matches the energy of the pairs in the plasma.
\end{abstract}
 
\maketitle
\section{Introduction}
In recent years, significant research has been devoted to quantum electrodynamics (QED) plasmas, driven by rapid advancements in high-intensity laser facilities \cite{DiPiazza}. These developments have enabled the exploration of various relativistic and nonlinear plasma phenomena, including the observation of pair creation at SLAC \cite{SLAC1997} and quantum radiation reaction at Rutherford \cite{Rutherford}. For a comprehensive review of recent progress in QED with intense background fields, see \cite{fedotov2023advances}.
Despite the steady progress in laser technology, the peak electric field achievable in laboratory settings remains well below the critical field $E_{cr}$ \cite{SLAC,LUXE}, defined as $E_{cr}=m^2c^3/e\hbar$, $e$ and $m$ are electron charge and mass and $\hbar$ is the reduced Planck constant. Nonetheless, several plasma-based schemes have been proposed to reach the critical field in laboratory environments \cite{Vinventi2021,Vincenti2024,PlasmaBasedFieldEnhancement2025}.

The process of electron-positron pair creation in vacuum via the Schwinger mechanism was first proposed by Sauter \cite{Sauter1931} and later developed by Schwinger \cite{Schwinger}. This mechanism has since been extensively studied in the literature. With the advancement of laser facilities over recent decades, increasing attention has been given to Schwinger pair creation. In particular, pair creation in vacuum using the Dirac-Heisenberg-Wigner (DHW) formalism has been studied \cite{Birula,sheng,Gies,schutzhold2008dynamically,Kohlfurst2013}.
As electron-positron pairs are continuously generated when strong fields propagate in vacuum, the resulting plasma current alters the original electromagnetic field, making back-reaction effects—mediated by Ampère’s law—significant. In \cite{QKT1,QKT2,QKT3,QKT4,QKT5,QKT6,QKT7,QKT8,QKT9,QKT10,QKT11,QKT12}, the plasma dynamics, initialized from both vacuum and pre-existing plasma states, were examined using quantum kinetic theory. Similarly, in \cite{PairAnnihilation2023}, the quantum Vlasov equation was employed to investigate plasma oscillations in electron-positron systems.
In \cite{2021,2023}, the pair creation including back-reaction has been studied using the DHW-formalism.
While the pair creation have been studied well, the pair annihilation remains less explored. 
The authors of \cite{PairAnnihilation2023} explored how pair creation are dependent on the properties of the initial plasma and presented some results on pair annihilation. They found that the annihilation process primarily increases the kinetic energy of the resulting particles, with no corresponding enhancement in the energy of the electric field.

In the present paper, we investigate collective pair annihilation in electron-positron plasmas. Our goal is to identify the conditions that favor annihilation over pair creation and to determine whether the energy released from annihilated pairs can enhance the amplitude of the electric field.
To this end, we adopt the DHW formalism and numerically study the evolution of time-dependent but spatially homogeneous fields in plasma. We find that plasma properties—such as temperature and density—strongly influence the likelihood of pair annihilation dominating over pair creation. Moreover, the energy released from annihilated pairs can enhance the electric field amplitude, provided that pair creation is sufficiently suppressed.
In contrast to \cite{PairAnnihilation2023} , which applies the Vlasov equation corrected by the Schwinger pair production rate—we employ a formalism based on the Wigner transformation of the Dirac Hamiltonian, allowing us to fully capture the relativistic quantum dynamics encoded in the Dirac equation.
This approach enables us to resolve the physics of high-frequency fields 
$\omega\sim \omega_c$ , where $\omega_c$
is the Compton frequency. At such frequencies, the polarization current arising from spin becomes comparable to—or even exceeds—the classical current, and must therefore be taken into account.
In addition to pair annihilation driven by strong field amplitudes (via the Schwinger mechanism), we also examine pair annihilation induced by high-frequency waves. The annihilation of high-energy photons leads to both creation annihilation of pairs at different momenta, resulting in either an increase or decrease in the total number of pairs. This outcome depends on both the initial plasma conditions and the frequency of the electromagnetic waves.

The organization of this paper is as follows: In \cref{Section2}, a brief description of the DHW formalism with the numerical system used in the work is given. In \cref{Section3}, the numerical results of the pair annihilation due to the Schwinger mechanism are presented. Then in \cref{Section4}, we have the results for the high-frequency pair annihilation. Finally, in \cref{Discussion}, a summary of the paper with conclusions is given.

\section{Basic equations}
\label{Section2}

The DHW formalism is a quantum kinetic theory of the Dirac field, first introduced in Ref. \cite{Birula}. Alternative derivations of the DHW formalism, with slight variations, have been presented in Refs. \cite{Gies,sheng}. A concise summary of the main steps in the derivation can also be found in, for example, Ref. \cite{2021}.
This quantum kinetic approach is based on a gauge-invariant Wigner transformation of the density matrix constructed from the Dirac four-spinors. The resulting theory is exact, aside from the Hartree-Fock (mean-field) approximation applied to the electromagnetic field.

The DHW equations consist of 16 coupled scalar phase-space variables, which describe mass (phase-space) density, charge density, current density, spin density, and other related quantities. However, when considering a one-dimensional electrostatic geometry with the electric field 
\(\mathbf{E} = E(z,t)\, \hat{\mathbf{z}}\), the full system (as detailed in Ref.~\cite{2021}) can be reduced to four coupled scalar equations, as follows:

\begin{align}
\label{PDE_System}
    D_t\chi_1(z,\bp,t)&= 2\varepsilon_{\bot} \chi_3(z,\bp,t)- \frac{\partial \chi_4}{\partial z} (z,\bp,t)\notag\\
    D_t\chi_2(z,\bp,t) &= -2p_z\chi_3(z,\bp,t)\\
    D_t\chi_3(z,\bp,t)&= -2\varepsilon_{\bot}(p_{\bot}) \chi_1(z,\bp,t) +2p_z\chi_2(z,\bp,t)\notag\\
    D_t\chi_4(z,\bp,t)&= -\frac{\partial \chi_1}{\partial z}(z,\bp,t)\notag 
\end{align}
together with the Ampère's law
\begin{equation}
\label{Ampers_law}
\frac{\partial E}{\partial t}=-\frac{e}{(2\pi)^3}\int \chi_1 d^3p
\end{equation}
where $D_t=\partial/\partial t +e{\tilde E}\partial/\partial p_z$ and $\varepsilon_{\perp}=\sqrt{m^2+p_{\perp}^2}$, and where $p_{\perp}$ is the perpendicular momentum and ${\tilde E}=\int_{-1/2}^{1/2} E(z+i\lambda\partial/\partial p_z,t)d\lambda$. The scalar variables $\chi_i$ (i=1,2,3,4) are dimensionless variables that have the following physical interpretation: Firstly, $\chi_1$ is the longitudinal component of the vector current density

\begin{equation}
    j_z=\frac{e}{(2\pi)^3}\int \chi_{1} d^3p
\end{equation}
where $j_z$ is the current.
The second variable $\chi_{2}$ represents the scalar mass density $\rho_m$, i.e.
\begin{equation}
    \rho_m =\frac{m}{(2\pi)^3}\int \frac{\chi_{2}}{\varepsilon_{\perp} } d^3p
\end{equation} 
Furthermore, $\chi_{3}$ can be considered as the spin density, i.e. the angular momentum density ${\bf M}$ due to the spin is 
\begin{equation}
    {\bf M} =\frac{1}{(2\pi)^3}\int ({\bf \hat{z}}\times {\bf p}) \frac{\chi_{3}}{2\varepsilon_{\perp} } d^3p
\end{equation} 
and finally $\chi_{4}$ represents the charge density $\rho_c$, i.e. 
\begin{equation}
    \rho_c=\frac{e}{(2\pi)^3}\int \chi_{4} d^3p.
\end{equation}
Note that the nonlocal electric field \( \tilde{E} \) reduces to the ordinary electric field \( E(z,t) \) when the spatial scale lengths are much larger than the characteristic de~Broglie wavelength. This simplification also holds naturally in the spatially homogeneous case. 
The system of equations is presented in natural units, where \( \hbar = c = 1 \).

We will now rewrite the system \cref{PDE_System} to simplify the numerical calculations. The first consideration when implementing a numerical solution to \cref{PDE_System} is the fact that \( \chi_1 \) and \( \chi_2 \) have nonzero vacuum contributions. These contributions arise from the expectation values of the free Dirac field operators (see, for example, Ref.~\cite{Birula,plasma_quantum_kinetic_review}), where the vacuum expressions are given by

  \begin{align}
 \chi_{1vac}&=-\frac{2 p_z}{\varepsilon}\notag \\ \chi_{2vac}&=-\frac{2\varepsilon_{\perp}}{\varepsilon}.
 \end{align}
 Here $\varepsilon=\sqrt{m^2+p^2}$. To avoid the problem of having variables with initial values that do not vanish at the boundaries, we introduce new variables $\Tilde{\chi}_i(z,\bp,t)$ as the deviation from the vacuum state, i.e. we let
 \begin{equation}
     \Tilde{\chi}_i(z,\bp,t)=\chi_i(z,\bp,t)- \chi_{i {\rm vac}}(\bp).
    \end{equation}
Note that \( \tilde{\chi}_{3,4} = \chi_{3,4} \). The new variables can now have zero values at the boundaries. 
A second modification to \cref{PDE_System} is the simplification of the operator \( D_t \). By switching to the canonical momentum, we simply get \( D_t = \frac{\partial}{\partial t} \). Using the Weyl gauge, where the scalar potential is zero such that \( E = -\frac{\partial A}{\partial t} \), the \( z \)-component of the kinetic momentum is replaced by \( q = p_z + eA \). In this gauge, the electric field is entirely described by the time derivative of the vector potential.
Our third modification involves switching to dimensionless variables (In the figures below, I used non-normalized variables to provide the reader with better intuition.). The normalized variables are given by:
\[
t_n = \omega_c t, \quad q_n = \frac{q}{mc}, \quad p_{n\perp} = \frac{p_{\perp}}{mc},
\newline
\quad E_n = \frac{E}{E_{\text{cr}}}, \quad A_n = \frac{eA}{mc},
\]
 Note that the DHW functions are already normalized. For notational convenience, we omit the index \( n \) and the tilde \( \tilde{} \) in the following.
The final modification to \cref{PDE_System} involves considering the homogeneous limit, which results in a variable charge density \( \chi_4 = 0 \). This reduces the phase-space scalars from 4 to 3. The equations to be solved numerically now read:

\begin{align}
    \frac{\partial  \chi_1}{\partial t} (q,p_{\bot},t)&= 2\varepsilon_{\bot} \chi_3 + 2E\frac{\varepsilon_{\bot}^2}{\varepsilon^3} \notag\\
      \frac{\partial  \chi_2}{\partial t} (q,p_{\bot},t )&= -2(q-A)\chi_3-2(q-A)E\frac{\varepsilon_{\bot}}{\varepsilon^3}\notag\\
      \frac{\partial \chi_3}{\partial t} (q,p_{\bot},t)  &=
      -2\varepsilon_{\bot} \chi_1 
      +2(q-A)\chi_2 
     \label{system}
      \end{align}
      with Ampère's law
      \begin{equation}
\label{Ampers_law2}
\frac{\partial E}{\partial t}=- \eta \int \chi_1 d^2p
\end{equation}
where we defined the dimensionless parameter \( \eta = \alpha / \pi \approx 2.322 \times 10^{-3} \), where \( \alpha \) is the fine-structure constant. This parameter determines the coupling between the electric field and the current density in natural units.
This system of equations describes the dynamics of an electron-positron plasma subjected to ultra-strong electric fields. The ions are treated as a stationary neutralizing background, and their motion is not included in the model.
Due to cylindrical symmetry, the azimuthal integration has already been performed, leading to the reduced phase-space volume element \( d^2p = p_{\perp}\, dq\, dp_{\perp} \).
The system of equations \cref{system}, together with Ampère's law \cref{Ampers_law2}, is solved numerically using a phase-corrected staggered leapfrog method~\cite{LeapFrog}. The simulation is carried out in three independent variables: time \( t \), longitudinal momentum \( q \), and perpendicular momentum \( p_{\perp} \). The typical discretizations parameters are \( \Delta t = 0.002 \), \( \Delta q = 0.01 \), and \( \Delta p_{\perp} = 0.1 \).
Although the problem is spatially homogeneous, a typical simulation involving strongly relativistic motion requires a large momentum cutoff \( q_{\text{max}} > \gamma_{\text{max}} \), where \( \gamma_{\text{max}} \) is the maximum relativistic Lorentz factor of the particles. As a result, the simulation becomes memory-intensive.
The accuracy of the numerical solution is verified through checks on energy conservation, based on the system’s underlying conservation laws.
 
\begin{align}
    &\frac{d}{dt}\bigg( \frac{E^2}{2}+ \eta\int d^2p \Big[\chi_2+ (q-A)\chi_1 \Big] \bigg)=0
\end{align}
for the DHW system and the Vlasov systems, respectively. Both of the conservation laws are fulfilled to a good approximation in the simulations, with relative numerical errors typically less than $10^{-4}$.

The DHW formalism is based on the Dirac equation. Let us briefly discuss the physical phenomena included in the coupled system of \cref{system} and \cref{Ampers_law2}. Due to the use of the mean-field approximation, the model does not account for single-particle processes such as Larmor emission, nor does it include few-particle interactions like Breit–Wheeler pair production.
The choice of a one-dimensional electrostatic field geometry ensures that both Breit–Wheeler pair production and radiation reaction effects are negligible compared to the dominant collective plasma dynamics. In this geometry, the electric field is longitudinal, meaning that the electric field in the rest frame of the electrons is the same as in the lab frame. This contrasts with a transverse field geometry, where the rest-frame electric field can be stronger than in the lab frame. As a result, the quantum parameter in the electrostatic case is given by $\chi = E / E_{cr}$, making collective effects more pronounced than radiation reaction or non-linear Breit–Wheeler processes, which require a significantly larger $\chi$ to become relevant.
For further details, including calculations that justify this assumption, see Ref.~\cite{2023}.
If all quantum effects are omitted from \cref{system} and \cref{Ampers_law2}, the model reduces to the 1D electrostatic limit of the relativistic Vlasov equations. For a more detailed comparison between the DHW formalism and the Vlasov approach, see \cite{2025DHWVSVlasov}.
However, quantum effects that manifest as collective phenomena are generally included in the DHW model. These include, for instance, collective pair creation (where the electric field is generated self-consistently by plasma currents), collective pair annihilation, Pauli blocking, spin polarization, and vacuum effects such as finite vacuum polarization~\cite{linear}.
Vacuum contributions also introduce the issue of charge renormalization. However, this effect is negligible when the simulation employs a momentum-space cutoff. For a detailed discussion, see Ref. \cite{2023}.

\section{Schwinger pair annihilation}
\label{Section3}
The annihilation of electron-positron pairs due to the amplitude of the electric field is the main focus of this section. 
\subsection{Vacuum}
\begin{figure}
    \centering
    \includegraphics[width=9 cm, height=10 cm]{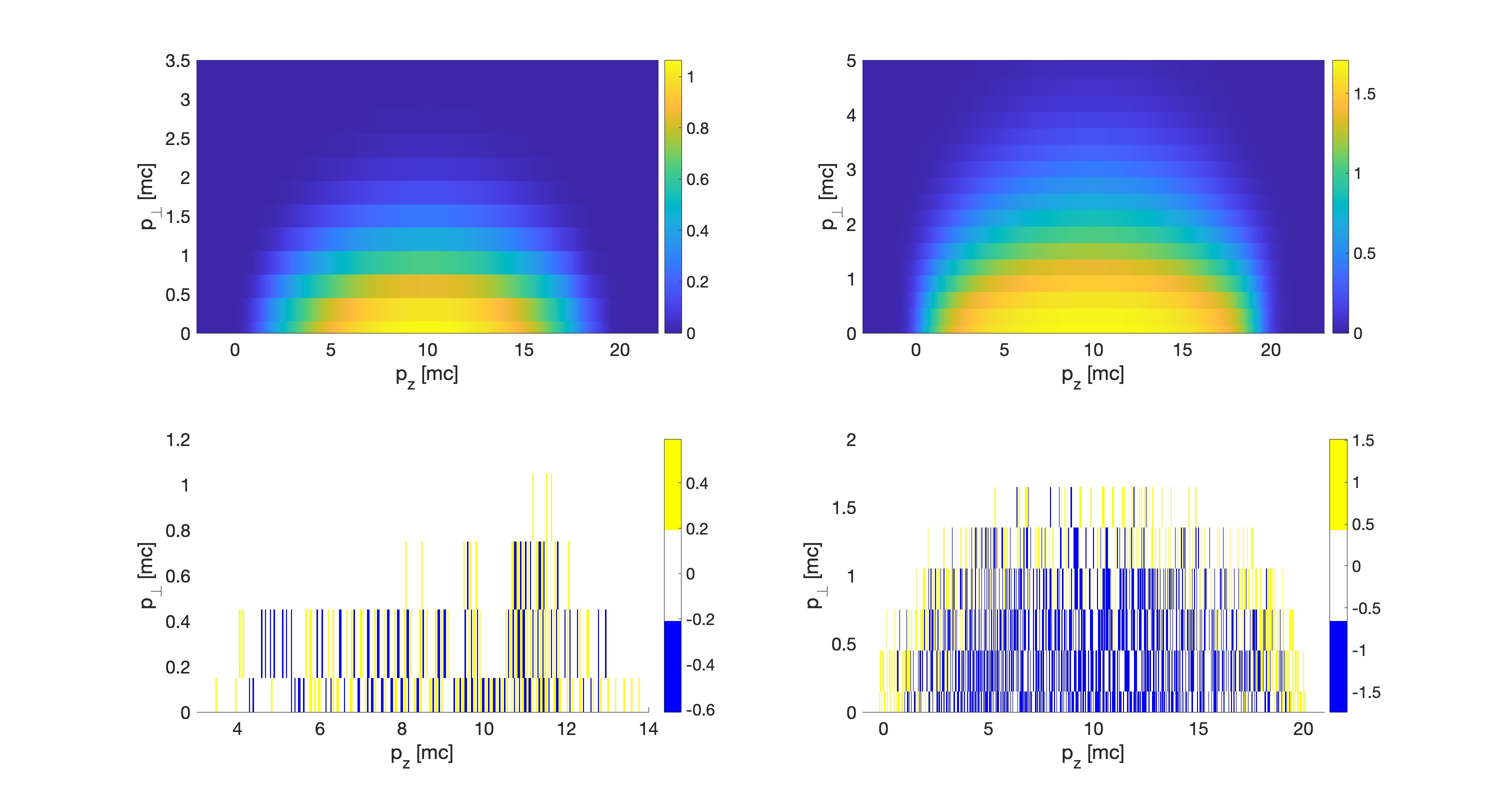}
    \caption{ The momentum distribution of the plasma that have been created by the two different Sauter pulses with $A_0=10$ and $t_0=10$. In the left column, we have used $\tau=2$ while in the right column we had $\tau=0.5$. For the upper row, we plot the plasma density at $t=10$ where the Sauter pulse has just vanished and we have the maximum pair density. In the lower row, we plot the pair density at $t=300$ minus the pair density at $t=10$. }
    \label{F(Q)}
\end{figure}

We start our numerical study by considering only vacuum initially (i.e., no plasma ) and use an external pulse that produces dense plasma. The applied external pulse is defined as 
\begin{align}
    A_{ex}(t) &=A_0 \bigg[1+ \tanh{\Big(\frac{t-t_0}{\tau}\bigg)}\Big]\\
    E_{ex}(t)&=-\frac{A_0}{\tau \cosh{\Big(\frac{t-t_0}{\tau}\Big)}^2}
\end{align}
The maximum amplitude of the electric field is given by \( A_0 / \tau \), while the peak value of the vector potential is \( 2A_0 \). In our simulations, we considered two cases with \( A_0 = 10 \) and \( t_0 = 10 \), allowing the pulse to gradually build up over time. We used two different pulse durations: \( \tau = 2 \) and \( \tau = 0.5 \).
The plasma is generated rapidly around \( t = 10\,\omega_c^{-1} \), by which time the Sauter pulse has essentially vanished. At this point, the resulting plasma densities are \( n = 4 \times 10^{30}\,\text{cm}^{-3} \) for \( \tau = 2 \) and \( n = 3 \times 10^{31}\,\text{cm}^{-3} \) for \( \tau = 0.5 \).
In \cref{F(Q)}, we plot the momentum distribution of the plasma for both cases. As expected, the pair density is higher at lower perpendicular momenta, since the probability of creating low-energy pairs is greater. 
In the upper-right panel of \cref{F(Q)}, the distribution shows significant Pauli blocking, with occupation numbers reaching up to 1.65 for lower perpendicular momentum states (the maximum allowed value is 2). The degree of blocking decreases with increasing perpendicular momentum, and for \( p_{\perp} \gtrsim 4 \), the states are nearly unoccupied.
In contrast, the upper-left panel shows lower occupation levels, with blocking up to 1 at small \( p_{\perp} \), and nearly free states for \( p_{\perp} > 2 \). Compared to the upper-right panel, where states are unoccupied only beyond \( p_{\perp} > 3.5 \), this indicates a less degenerate plasma in the \( \tau = 2 \) case.

After the Sauter pulse has vanished, the system undergoes plasma oscillations with peak electric field amplitudes of \( E_0 = 0.75\,E_{\text{cr}} \) and \( E_0 = 1.96\,E_{\text{cr}} \), respectively. These oscillations correspond to a peak vector potential of \( A_0 = 20.1\,mc \) in both cases. The strong electric fields during the plasma oscillation are sufficient to trigger both pair creation and annihilation processes.
In the lower-left panel of \cref{F(Q)}, we observe that in the lowest perpendicular momentum states, both pair creation and annihilation take place, with the two processes nearly offsetting each other. Positive values indicate newly created pairs, while negative values correspond to annihilated pairs. The distribution reaches up to a value of 2, representing a fully occupied energy state for electron-positron pairs.
By integrating over momentum space, we find that the total number of particles remains approximately conserved. For higher perpendicular momenta, the net effect of pair production and annihilation diminishes. This behavior is expected, as the Schwinger pair production probability decreases with increasing perpendicular momentum due to the associated increase in effective mass, which in turn reduces the exponential rate.

In the lower right panel of \cref{F(Q)}, we observe that for $p_{\perp}<1$ and $p_z<20$, more electron-positron pairs are annihilated than created. By integrating over momentum, we find that the total number of particles has decreased. This can be attributed to the dense plasma generated by the Sauter pulse, which occupies a significant region of longitudinal momentum space (up to approximately $24 mc$). The value of $F(q)$ reaches around $1.6$, indicating that most energy states are occupied, leaving few vacant states (with $2$ being the maximum possible due to spin degeneracy).
Although the electric field reaches a relatively high amplitude ($1.96,E_{cr}$), which would typically favor pair creation, the high occupation of low-energy states suppresses this process due to Pauli blocking. Furthermore, because the plasma undergoes longitudinal motion due to plasma oscillations, it shifts in momentum space. However, since the vector potential amplitude $A_0 = 20.1$ is smaller than the width of $F(q, p_{\perp}=0)$, the creation of new low-energy pairs is strongly suppressed; only high-energy pairs can be generated. Consequently, the plasma oscillation tends to favor annihilation over creation in this regime. Note that the plasma shown in the left column had a momentum width smaller than $A_0$, making it possible to create electron-positron pairs. For higher perpendicular momenta, $1 < p_{\perp} < 1.5$, we observe that pair creation dominates over annihilation. This occurs because Pauli blocking becomes less significant at higher perpendicular momentum, where fewer states are initially occupied.

\begin{figure}
    \centering
    \includegraphics[width=9 cm, height=10 cm]{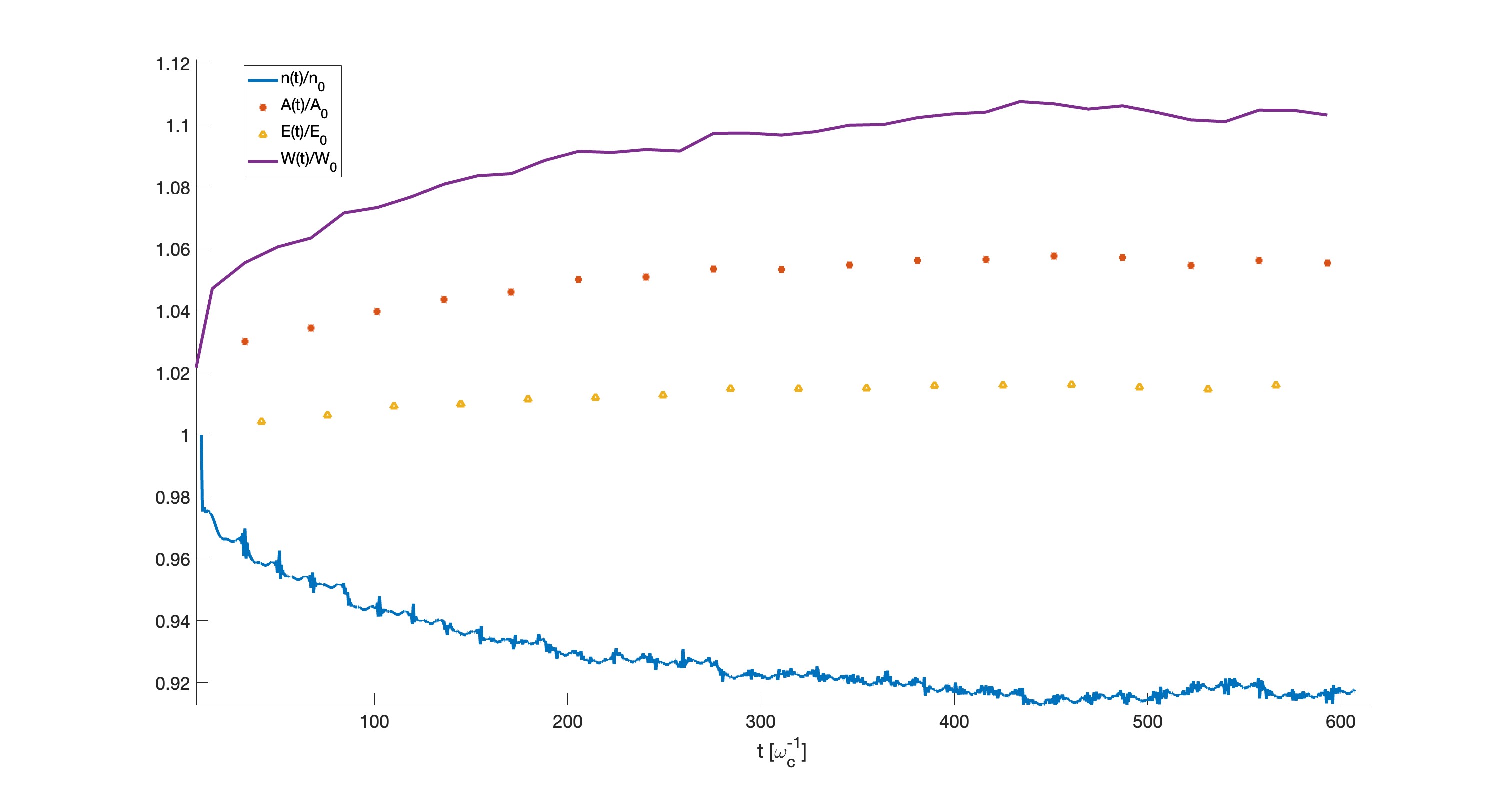}
    \caption{The relative change of pair density $n(t)/n_0$ (dashed curve), the relative change of the vector potential $A(t)/A_0$ (pluses), the relative change of the field amplitude $E(t)/E_0$ (triangles) and the relative change of the energy per particle $W(t)/W_0$ (solid line) vs. time. }
    \label{AnnhilationVacuum}
\end{figure}
 
A longer simulation corresponding to the right panel of \cref{F(Q)} is analyzed in further detail in \cref{AnnhilationVacuum}. Approximately 8 $\%$  of the plasma has annihilated by this point; this fraction would be higher if the creation of new pairs were blocked. At later times, the processes of pair creation and annihilation become roughly equally probable. The energy released from the annihilated pairs is transferred to the remaining plasma particles, resulting in an increase of about 11$\%$ in the energy per particle, as shown in \cref{AnnhilationVacuum}. Additionally, the vector potential—which represents the average relativistic gamma factor during plasma oscillations—increases by approximately $5\%$ due to the redistribution of energy from the annihilated pairs. This enhancement of the vector potential subsequently strengthens the electric field, which is fundamental for plasma oscillations.  Consequently, the amplitude of the electric field increases by approximately $1.6\%$. In contrast to previous studies \cite{PairAnnihilation2023}, our results demonstrate that pair annihilation can lead to an enhancement of the electric field amplitude.

\subsection{Initial Plasma}
In this subsection, we begin our simulation with a preexisting plasma. From the previous subsection, we have learned that the plasma distribution significantly affects the probability of pair annihilation. Specifically, when the plasma distribution is larger than 1, it becomes more probable for the pairs to be annihilated rather than for new pairs to be created. In DHW theory, the background plasma and the vacuum both contribute to the mass and current, as expressed by the equation
\begin{equation}
F = f_e + f_p - 1
\end{equation}
where $f_e$ and $f_p$ represent the electron and positron densities, respectively, and $-1$ accounts for the vacuum state. If the densities of electrons and positrons are low and/or their energies are high, the number of electrons and positrons per energy state is less than 1, which results in $F < 0$. This implies that the probability of pair annihilation becomes negligible. Consequently, in order to observe more pair annihilation than pair creation, it is necessary to start with a plasma that has high density and low temperature such that we get $F>0$.
Another key observation from the previous section is that more pair annihilation than creation can only occur if the momentum spread of the plasma is larger than the vector potential $A_0$. Therefore, we must focus on studying plasmas that satisfy this condition.

As an initial condition for the plasma, we use a Fermi-Dirac distribution $f_{FE}$ representing both electrons and positron
\begin{equation}
    f_{FE}=f_e+f_p=\frac{2}{1+e^{(\varepsilon-\mu_n)}/T_n}
\end{equation}
where \( \mu_n \) and \( T_n \) are the normalized chemical potential and temperature, respectively. Note that the factor of 2 arises from the contribution of both electrons and positrons. Consequently, we can have \( F_{\text{max}} = 1 \), representing fully occupied energy states.  Note that this distribution is not in thermodynamical equilibrium. The aim is not to begin with a state of thermodynamic equilibrium, but rather to use one that shares a similar structure with the distribution generated by the Sauter pulse in the previous subsection. Starting with a thermally equilibrated distribution would result in negligible collective annihilation.
Using this representation of the plasma in the DHW system, we obtain:
\begin{align}
    \chi_1&=\frac{2 q f_{FE}}{\sqrt{1+q^2+p_{\perp}^2}}\\
    \chi_2&=\frac{2 \sqrt{1+p_{\perp}^2} f_{FE}}{\sqrt{1+q^2+p_{\perp}^2}}\\
    \chi_3&=0
\end{align}
where the last relation corresponds to the absence of spin polarization in the initial distribution. To obtain a plasma distribution with a wide momentum spread that fills all low-energy states, a large \( \mu_n \) is required. Moreover, the plasma must be sufficiently dense so that the vector potential \( A_0 \) remains relatively small; otherwise, the plasma would be accelerated to high energies, reopening the possibility of low-energy pair creation. One way to ensure \( A_0 \) is smaller than the momentum spread is to use a weak field amplitude for the plasma oscillation, but this would suppress the Schwinger pair production probability. Therefore, a field amplitude must be chosen that enhances the Schwinger mechanism while simultaneously keeping \( A_0 \) low enough to avoid excessive plasma acceleration.

\begin{figure}
    \centering
    \includegraphics[width=9 cm, height=10 cm]{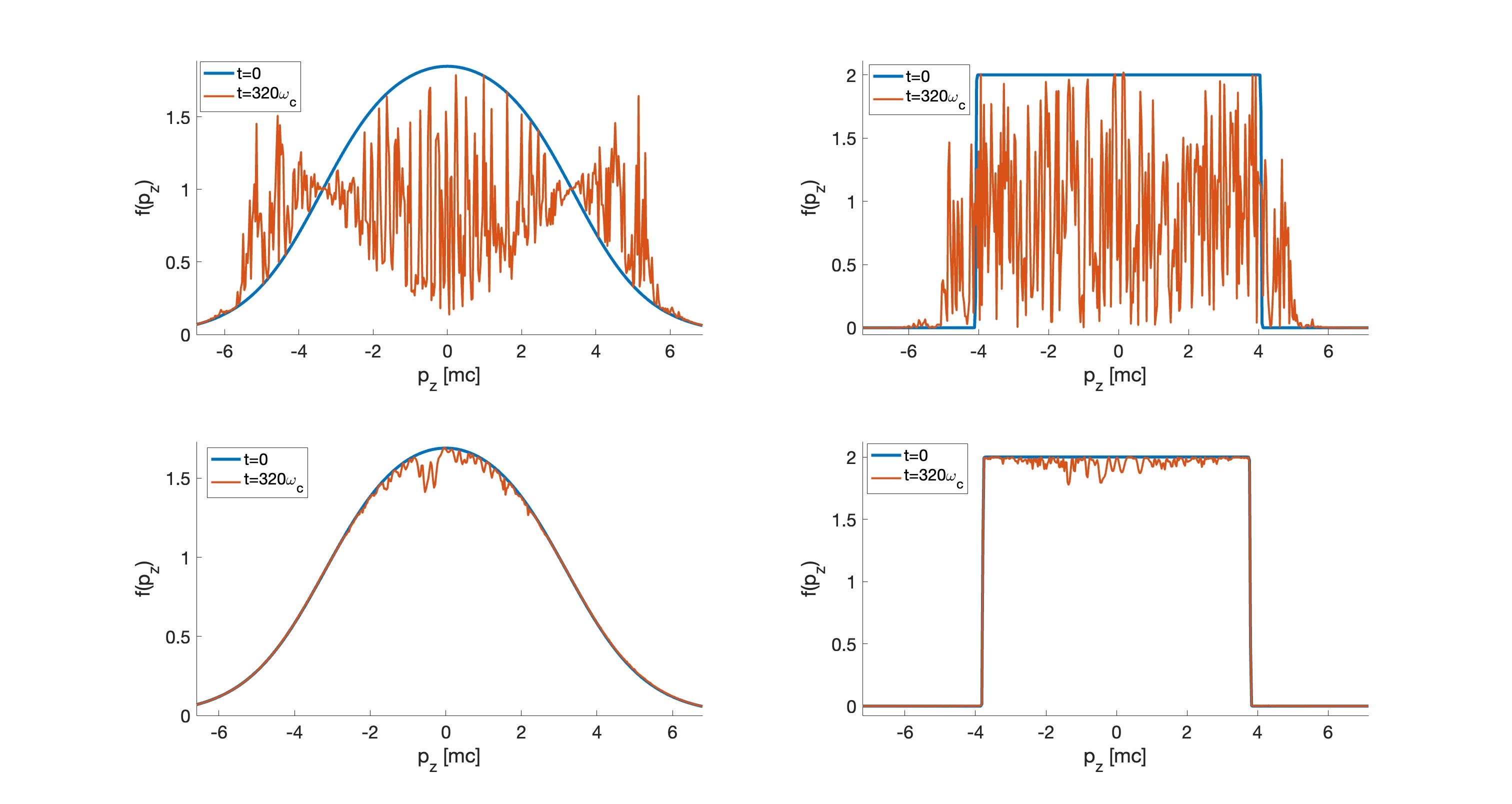}
    \caption{The parallel momentum distribution \( f(p_z) \) of the plasma is plotted at \( t = 0 \) (solid curve) and \( t = 320 \) (dashed curve). In the left column, the plasma has initial parameters \( \mu = 3.5 \) and \( T = 1 \), while in the right column, \( \mu = 4.2 \) and \( T = 0.005 \). The upper row shows the distribution at the lowest perpendicular momentum \( p_{\perp} = 0 \), whereas the lower row corresponds to \( p_{\perp} = 1.5 \).}
    \label{TempFigure}
\end{figure}

We begin by considering an electron-positron plasma with \( \mu_n = 3.5 \) and \( T_n = 1 \), which corresponds to a plasma density of \( n = 4 \times 10^{31} \, \text{cm}^{-3} \). The plasma oscillation is initiated by applying an electric field with an initial amplitude of \( E(t = 0) = 1.5 E_{\text{cr}} \). This results in a vector potential of \( A_0 = 5.7 \, \text{mc} \).

In the left column of \cref{TempFigure}, we plot the initial plasma distribution alongside the distribution at approximately \( t = 320 \omega_c \), at which point around \( 2\% \) of the initial plasma has been annihilated. In the first row of \cref{TempFigure}, we show the plasma momentum distribution at the lowest perpendicular momentum, while in the lower row, we examine \( p_{\perp} = 1.5 \). Both created and annihilated pairs are more prominent at lower perpendicular momentum because the Schwinger probability for both processes is higher in this region. For higher perpendicular momentum, there are clearly fewer particles that have been annihilated, and almost no new particles have been created. This is because both longitudinal and perpendicular momenta are large in this region, which suppresses pair creation, whereas pair annihilation only requires a high perpendicular momentum. The amount of newly created pairs relative to the initial plasma is \( 0.5\% \), while the percentage of annihilated pairs is \( 2.5\% \), resulting in a net decrease of \( 2\% \). A closer inspection of the upper-left figure in \cref{TempFigure} reveals that most of the new pairs are created along the curve of the initial plasma that was already non-zero. This indicates that the newly created pairs have filled energy states that were not completely empty.
If we were to reduce the temperature, we could redistribute the electron-positron plasma in such a way that all energy states are fully occupied. However, if we kept the same value of \( \mu_n \) and set \( T_n = 0 \), the plasma density would decrease. To maintain the same plasma density, we would need to increase the chemical potential to \( \mu_n = 4.2 \). Using the same initial electric field amplitude, we plotted the plasma distribution function in the second column of \cref{TempFigure}. This shows a sharper distribution cutoff at \( 4 \, \text{mc} \), compared to \( 6 \, \text{mc} \) in the left column. Despite this, the number of pairs annihilated is higher because the combined electron-positron plasma and vacuum states result in \( F = 1 \) over a larger momentum range. The percentage of annihilated pairs relative to the initial plasma is \( 4.25\% \), while the percentage of created pairs is \( 0.3\% \). The number of pairs created is slightly lower because the initial plasma has a sharper distribution, preventing totally pair creation within the plasma box, compared to the hotter plasma that allowed more pair creation. In the lower-right panel of \cref{TempFigure}, pair annihilation is also weak for the same reason as in the lower-left panel.

\begin{figure}
    \centering
    \includegraphics[width=9 cm, height=10 cm]{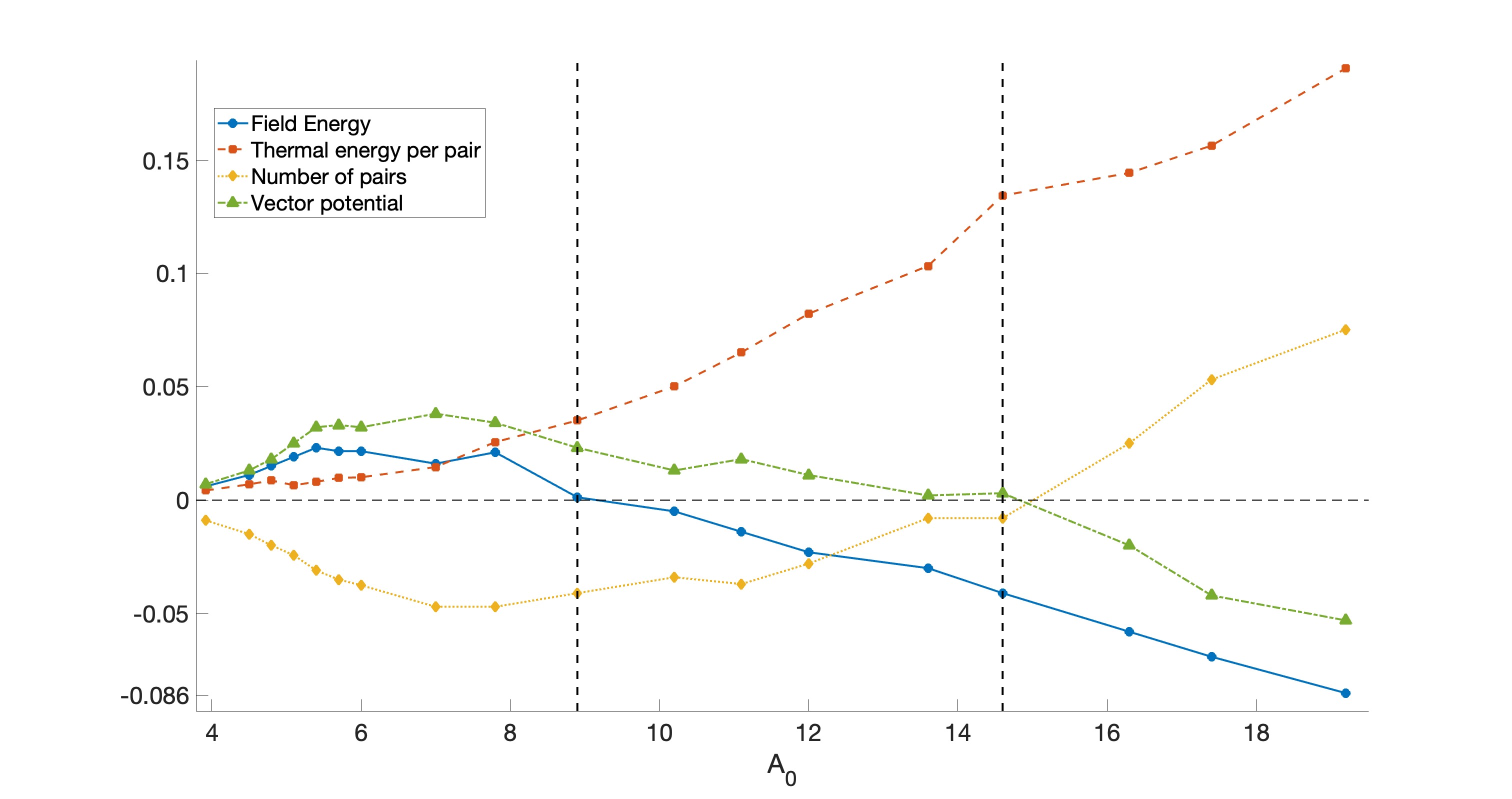}
    \caption{A plot showing the relative change in several physical quantities after five plasma oscillations is presented. The relative change is defined as the difference between the value at the fifth plasma oscillation and the initial value, divided by the initial value, and is plotted against the vector potential $A_0$. The solid line with circles represents the field energy, the dashed line with circles corresponds to the thermal energy per pair, the dotted line with triangles denotes the number of pairs, and the dashed line with triangles indicates the vector potential. }
    \label{EnergyPlot}
\end{figure}

As the pair density in momentum space changes due to pair creation and annihilation, both the electric field energy and the particle energy are affected. In the previous subsection, we demonstrated that the energy per pair and the field energy increase as the number of pairs decreases due to pair annihilation. In this subsection, we investigate further how the different energy components of the plasma and the field are impacted by changes in the number of pairs. We consider a plasma with a Fermi-Dirac distribution, with \( T = 0.005 \) and \( \mu = 5.2 \), and explore a range of different values by varying the field amplitude \( E_0 \).
As the field and particle energies oscillate during plasma oscillations, we focus on how their peak values evolve over time. For the electric field energy, we evaluate the moment when the plasma particles attain their minimum energy and the field reaches its peak energy. For the energy per particle, as previously defined, we can calculate it as the peak of the total particle energy divided by the number of pairs. Alternatively, by evaluating the particle energy at the instant when the field is maximized, corresponding to \( A_0 = 0 \), we obtain the thermal energy of the plasma. This approach allows us to study how thermal energy responds to variations in pair density. Meanwhile, the evolution of total particle energy can still be monitored via the vector potential \( A_0 \), which represents the average relativistic gamma factor over an oscillation cycle. In \cref{EnergyPlot}, we show the relative changes in field energy, thermal energy per pair, pair number, and vector potential, all normalized to their initial values. These quantities are evaluated after five plasma oscillations and plotted as functions of the vector potential \( A_0 \).

In the low-\(A_0\) region, the low-temperature plasma is accelerated to low momentum while still blocking the low-energy states, thereby preventing the creation of low-energy pairs. As a result, pair annihilation becomes the dominant process, as evident from the negative values of the relative change in the number of pairs. The energy released from the annihilated pairs increases the kinetic energy of the remaining plasma, which can be clearly observed by tracking the relative change in the vector potential \(A_0\). A decrease in the number of pairs leads to an increase in \(A_0\), while an increase in the number of pairs leads to a decrease in \(A_0\).
In the low-\(A_0\) region, where pair annihilation is the dominant process, an increase in \(A_0\) leads to a rise in the field energy. Regarding the thermal energy per pair, pair annihilation reduces the total thermal energy, but the reduction in the number of pairs is more significant, resulting in an overall increase in thermal energy per particle. This increase continues gradually until the reduction in pair number reaches its maximum around \(A_0 = 7\). For \(A_0 > 7\), the plasma is accelerated to momenta sufficiently large compared to the initial momentum width of the plasma (\(\sim 5.2\)), allowing the creation of low-energy pairs. This reduces the rate of pair annihilation and increases the total thermal energy of the plasma, as the newly created pairs extend the momentum distribution of the initial plasma.

This is confirmed by the steeper rise of the relative change in thermal energy per pair for \(A_0 > 7\). The creation of new pairs requires energy, which is extracted from the field energy, thereby slowing the growth rate of the latter. At \(A_0 = 9\), the growth of the field energy effectively halts. In the region \(9 < A_0 < 15\), pair annihilation continues, but the field energy begins to decrease. This occurs because the initial plasma is now accelerated to higher momenta, and the low-energy states remain accessible for a longer portion of the plasma oscillation cycle, enabling the creation of a significant number of new pairs. However, the total number of pairs still decreases, as the annihilation rate exceeds the rate of pair creation. Consequently, although pair annihilation contributes to the field energy, the energy consumed in creating new pairs outweighs this gain, leading to a net decrease in the field energy. Meanwhile, the vector potential, which is related to the number of pairs in the system, continues to increase as long as the number of pairs is decreasing.

For \(A_0 > 15\), pair creation becomes the dominant process over pair annihilation, as the initial plasma is accelerated to momenta exceeding approximately three times its momentum width. Both the field energy and the vector potential are reduced compared to their initial values. Although the total number of pairs increases beyond the initial count, the thermal energy per pair also rises due to the broader momentum distribution caused by the creation of new pairs, which enhances the overall thermal spread of the plasma.

\section{High-frequency pair annihilation}
\label{Section4}
In this section, we focus on pair annihilation induced by high-frequency waves in the plasma. For waves with a frequency \(\omega > 2\omega_c\), pairs can be created from the vacuum, see \cite{linear} for more details. The question arises: can high-frequency waves annihilate electron-positron pairs? To explore this, we apply the same principles used in the previous section for Schwinger pair annihilation, where we block all low-energy states, but now consider the effects of high-frequency waves.

To study pair annihilation induced by high-frequency waves, we begin with a plasma that satisfies the Fermi-Dirac distribution. To ensure the plasma remains non-degenerate, we choose parameters such that \(T = 0.001\) and \(\mu = 1.1\). We do not use a large chemical potential (\(\mu\)) in this case, as the effect is driven by high-frequency waves rather than a strong field amplitude. This implies that the vector potential is very small, i.e., \(A_0 \ll 1\). The high-frequency waves propagating through the plasma will create or annihilate pairs, depending on the availability of vacant energy states, with the energy \(\hbar \omega\).
For frequencies higher than $2\omega_c$ (more energy than needed to create a real pair), the additional energy is either transferred to the created pairs in the case of pair creation, or to the plasma and the wave amplitude in the case of pair annihilation. For a plasma with the initial properties described above, all energy states are blocked up to $\varepsilon \sim 0.5$. Therefore, considering the frequency range $2\omega_c < \omega < 2.5\omega_c$ (though this is not strict, as the probability is more fluid), pair creation should be blocked. This plasma will suppress almost all pair creation up to $\varepsilon \sim 0.5$. We now consider the following external pulse:

\begin{equation}
    E_{ex}=E_0 \sum_{i}\sin{(\omega_i t) }\,e^{-(t-t_{0i})^2/\tau^2}
\end{equation}
where we sum over different frequencies $\omega_i$ that are introduced to the plasma at different times $t_{0i}$, and $\tau$ is the time it takes for the pulse to decay to $1/e$ of its peak value. We set $E_0 = 0.001$ to suppress the probability of the Schwinger mechanism and use $\tau = 100$. Since the minimum frequency required to create or annihilate pairs is $2 \omega_c$, we set the frequency of the first pulse package as $\omega_1 = 2 \omega_c$ with $t_{01} = 0$. The initial plasma can block up to $\varepsilon = 0.5$, so a wave that delivers photons with slightly higher energy than the rest mass would increase the probability of pair annihilation. This is because photons with $\omega = 2 \omega_c$ will target the lowest energy states of the plasma, leaving the higher energy pairs of the initial plasma unchanged. As the second pulse package, we set $\omega_2 = 2.1 \omega_c$ and $t_{02} = 800$, providing the plasma with a possibility of a break from exposure to external pulses.

\begin{figure}
    \centering
    \includegraphics[width=9 cm, height=10 cm]{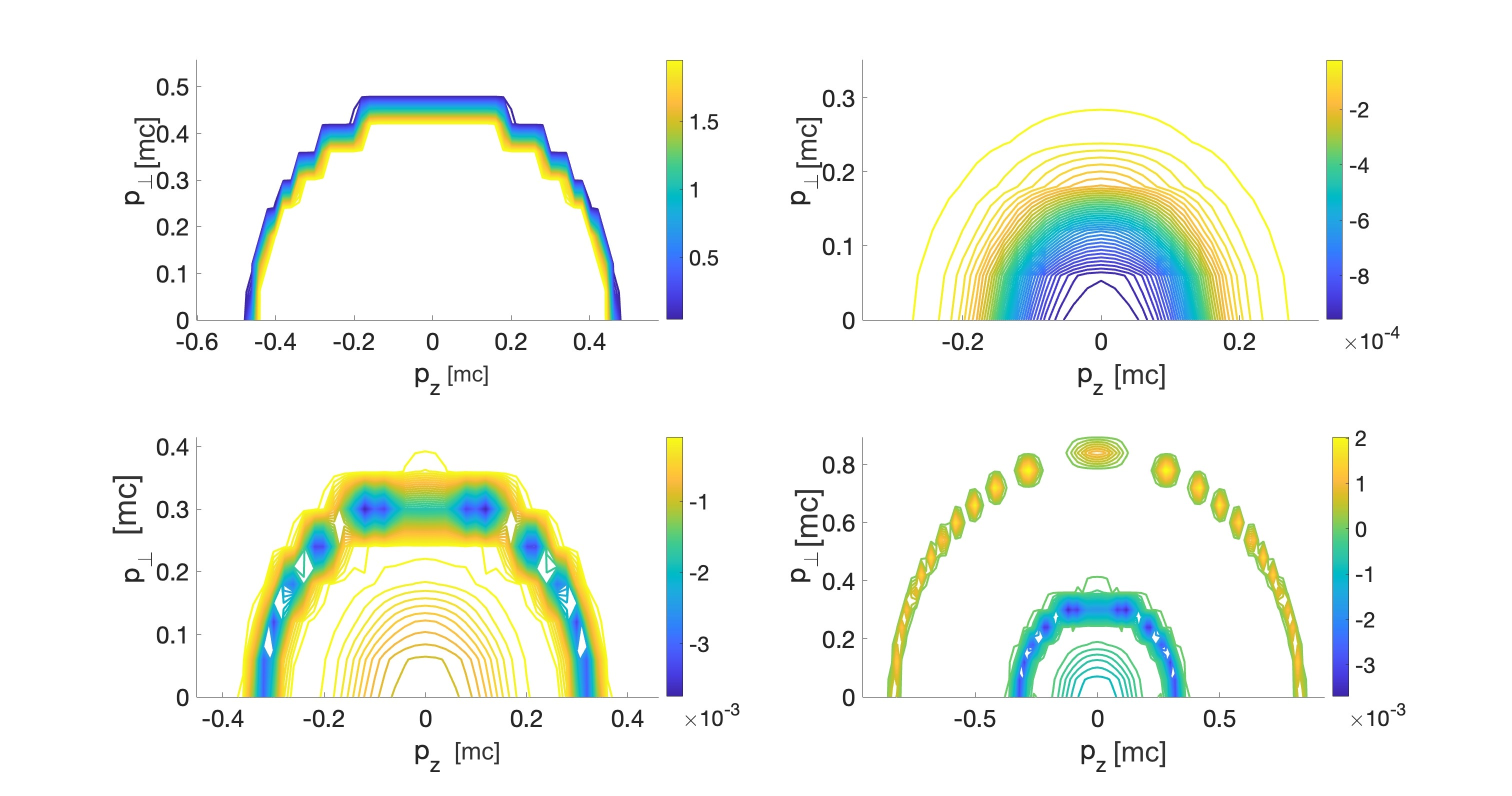}
    \caption{The momentum distribution of the electron-positron plasma is shown in the following panels. In the upper left panel, we display the initial plasma with $\mu = 1.1$ and $T = 0.001$. In the upper right panel, we show the difference in the pair plasma distribution at $n(t=200)$ (right after the first wave) minus the initial distribution $n_0$. In the lower left panel, we present the difference in the pair plasma distribution at $n(t=1200)$ (right after the second wave) minus the initial distribution $n_0$. Finally, in the lower right panel, we show the difference of $n(t=2000)$ after the third wave and the initial distribution $n_0$.  }
    \label{PairDistribution}
\end{figure}
As a third wave package, we consider a wave with photons that have higher energy than the pairs in the plasma, which leads to pair creation rather than the annihilation of real ones. We set $\omega_3 = 2.6 \omega_c$ and $t_{03} = 1600$. The momentum distribution of the created and annihilated pairs is presented in \cref{PairDistribution}. In the upper left panel of \cref{PairDistribution}, the initial distribution of the plasma is shown. As can be seen, all energies less than $\varepsilon = 0.5$ are blocked.
In the upper right panel, we plot the difference between the plasma distribution at $t = 200$ and the initial distribution, i.e., $n(t=200) - n(t=0)$, right after the first pulse package with $\omega = 2 \omega_c$ was delivered. As can be seen, only negative values are displayed, representing the pairs that have been annihilated. The photons annihilated pairs with energies up to $\varepsilon = 0.25$, but most of the annihilated pairs had very low energy in addition to the rest mass energy. The amount of pairs annihilated is $0.002\%$ of the initial plasma. This ratio is very small, but not surprisingly so, as the effect is linear and the initial plasma was dense with $n_0 = 5 \times 10^{28}/\text{cm}^3$.

For the left lower panel of \cref{PairDistribution}, we plot the momentum distribution difference $n(t=1200) - n(t=0)$ right after the second wave with $\omega = 2.1 \omega_c$ has vanished. The annihilated pairs are concentrated around $\varepsilon \sim 0.35$ as the photons have delivered more than just rest energy. The number of annihilated pairs is $0.025\%$, approximately 12 times more pairs than the first wave.
For the right lower panel, we plot the momentum distribution of $n(t=2000) - n(t=0)$. Here, the third wave with $\omega = 2.6 \omega_c$ has created new pairs in the system, which are concentrated around $\varepsilon = 0.8$. The third wave created around $0.038\% n_0$ of new pairs.

\section{Discussion}
\label{Discussion}
The main objective of this study is to investigate pair annihilation and to identify the conditions that favor it over pair creation. It is observed that in the presence of a strong electric field, \( E \sim E_{\text{cr}} \), both pair creation and annihilation occur. Starting from vacuum, it is found that pair annihilation is possible if the sum of the contributions from electrons, \( f_e \), and positrons, \( f_p \), exceeds the vacuum baseline of \(-1\); that is, if \( F > 0 \). This is confirmed by initializing a low-temperature plasma that satisfies this condition, where pair annihilation is indeed observed.
To maximize the number of annihilated pairs, the pair annihilation rate must be enhanced. This is achieved by initializing the electron-positron plasma such that it occupies all low-energy states (\( F = 1 \)) with low perpendicular momentum, \( p_{\perp} \leq \sqrt{E} \). For larger perpendicular momenta, \( p_{\perp} \gg 1 \), the annihilation rate becomes very low. Regarding the parallel momentum \( p_z \), the plasma is accelerated back and forth along the \( z \)-direction. Thus, a plasma with a broad parallel momentum distribution can still experience annihilation at peak probability.
Another factor to consider for maximizing pair annihilation is suppressing the pair creation process. This can be achieved by starting from a very low-temperature but high-density plasma with a parallel momentum spread larger than the plasma oscillation–averaged momentum, \( A_0 \). In this way, the low-energy states remain occupied by the initial plasma, effectively blocking pair creation.

The impact of pair creation and annihilation on the energy of both the plasma and the field has been investigated. As long as pair creation is suppressed and pair annihilation is occurring, the field energy increases. This is because the energy released from the annihilated pairs contributes to the kinetic energy of the remaining plasma, leading to an increase in the vector potential . As a result, the electric field amplitude is enhanced when pair annihilation dominates and pair creation is effectively blocked by the plasma.
Once the plasma is accelerated to higher momenta, allowing low-energy states to become available, pair creation resumes. At this stage, two competing processes influence the field energy: one enhances it through pair annihilation, and the other reduces it via pair creation. The latter has a stronger impact, leading to a net decrease in field energy—even if the total number of pairs continues to decrease.
This behavior has been reported in \cite{PairAnnihilation2023}, where the author concluded that pair annihilation does not lead to an increase in field energy. While this statement may hold under specific conditions, it is important to note that pair annihilation an enhance the field energy if the plasma effectively blocks the low-energy states.
Regarding the vector potential, as long as the total number of pairs is decreasing, the vector potential  continues to increase.

Another effect investigated in this work is pair annihilation induced by high-frequency waves. It is shown that waves with frequencies \(\omega \geq 2\omega_c\), when propagating through low-temperature plasma, can lead to both pair annihilation and creation. If the wave frequency corresponds to the energy of existing pairs in the plasma, resonance occurs, resulting in the annihilation of pairs. On the other hand, if the wave frequency exceeds the energy of the plasma pairs, it can lead to the creation of new pairs.

This work has been conducted using an electrostatic field geometry. For future studies, investigating pair annihilation in the presence of a full electromagnetic field in plasma is of interest. In such a scenario, the plasma is accelerated in multiple directions, resulting in richer and more complex physical dynamics.

\section{acknowledgment}
The author acknowledges support from the Knut and Alice Wallenberg Foundation. A special thank to Gert Brodin and David Reis for fruitful discussions.
 \bibliography{refs}

\end{document}